\documentclass[usegraphicx,useAMS]{mn2e}

\usepackage{amssymb}  
\usepackage{epsfig}

\def\eps@scaling{1.0}%
\newcommand\epsscale[1]{\gdef\eps@scaling{#1}}%

\newcommand\plotone[1]{%
 \centering
 \leavevmode
 \includegraphics[width={\eps@scaling\columnwidth}]{#1}%
}%

\newcommand\plottwo[2]{%
 \centering
 \leavevmode
 \columnwidth=.48\textwidth
 \includegraphics[width={\eps@scaling\columnwidth}]{#1}%
 \hfil
 \includegraphics[width={\eps@scaling\columnwidth}]{#2}%
}%

\title[]{On the incidence of chemically peculiar stars in the Large Magellanic cloud}
\author[Paunzen et al.]
       {E.~Paunzen,$^{1}$ O.I.~Pintado,$^2$\thanks
       {Member of Carrera del Investigador del Consejo Nacional de 
        Investigaciones Cient\'{i}ficas y T\'{e}cnicas de la Rep\'{u}blica 
        Argentina and Visiting Astronomer at Complejo Astron\'omico El Leoncito 
        operated under agreement between Consejo Nacional de Investigaciones 
        Cient\'{i}ficas y T\'{e}cnicas de la Rep\'ublica Argentina and the 
        National Universities of La Plata, C\'ordoba y San Juan} H.M.~Maitzen$^1$
		and A.~Claret$^3$
        \\       
        $^1$Institut f\"ur Astronomie, Universit\"at Wien,
        T\"urkenschanzstra{\ss}e 17, A-1180 Wien, Austria
        (Ernst.Paunzen@univie.ac.at) \\       
	    $^2$Departamento de F\'isica, Facultad de Ciencias Exactas 
        y Tecnolog\'ia, Universidad Nacional de Tucum\'an, Argentina \\ Consejo Nacional 
        de Investigaciones Cient\'ificas y T\'ecnicas de la Rep\'ublica Argentina \\
		$^3$Instituto de Astrof\'{\i}sica de Andaluc\'{\i}a
		CSIC, Apartado 3004, 18080 Granada, Spain
}

\date{Submitted 2005}

\pagerange{\pageref{firstpage}--\pageref{lastpage}}
\pubyear{2005}

\begin{document}

\maketitle

\label{firstpage}

\begin{abstract}
With the aim to corroborate the result of a search for chemically peculiar stars in the 
Large Magellanic Cloud (LMC) we present measurements obtained from CCD-imaging of two 
fields, one containing a young open cluster (NGC 1711). While for the latter field, 
including its surrounding we obtain a contribution of three percent of chemically 
peculiar stars detectable by $\Delta a$ photometry (i.e. the magnetic objects of
this group), the second field yields about half of this value in good accordance with 
the finding for NGC 1866 (Maitzen et al. 2001) the surrounding field of which has been 
found to exhibit a very low value of such stars - 0.3\%. Thus we are faced with the fact, 
that our incipient impression about a substantially lower appearance of magnetic 
chemically peculiar stars in the LMC as compared to the Galaxy continues to be valid.
Most of the photometrically identified peculiar stars (from their historical origin 
denominated Ap-stars) are located in the domain of the B-type stars. But this is  
a selection effect due to the limiting magnitude of our observing  conditions impeding 
the observation of fainter main sequence stars. In addition to objects showing up as 
positive deviators in $\Delta a$ photometry we also discuss nine stars which appear opposite the 
main line of normal stars, hence are negative deviators. For most of them the
interpretation as emission stars of B-type seems to be appropriate.
The statistically relevant number of observations obtained so far in the LMC supports 
the view that the formation of magnetic peculiar stars has occurred there at a significantly 
lower rate.
\end{abstract}


\section{Introduction}   \label{Sect1}

Since our pioneer work in which we have reported about the detection
of the first extragalactical chemically peculiar (CP) stars (Maitzen, Paunzen
\& Pintado 2001), a huge effort was spent in order to make follow
up observations in the Large Magellanic Cloud (LMC hereafter).
Such measurements are rather time consuming when using a narrow band
photometric system such as the three filter $\Delta a$ CCD system, but
much more efficient than spectroscopic observations. 

We have focussed our efforts on the incidence of chemically peculiar stars
in the LMC using the $\Delta a$ photometric system (Maitzen 1976). It measures the 
characteristic broad band absorption feature located around 5200\AA\
first described by Kodaira (1969). This flux depression is most certainly a consequence 
of the non-solar
elemental abundance of CP and related objects in the presence of a strong stellar
magnetic field (Kupka et al. 2004). The $\Delta a$ photometric system measures the flux 
depression at 5200\AA\ 
by sampling the depth of it, comparing the flux at the center (5205\AA, $g_{\rm 2}$),
with the adjacent regions (5030\AA, $g_{\rm 1}$ and 5510\AA, $y$) using bandwidths from
110\AA\ to 220\AA. The respective index $a$ was introduced as
$$ a = g_{\rm 2} - (g_{\rm 1} + y)/2 $$
Since this quantity is slightly dependent on temperature (increasing
towards lower temperatures), the intrinsic peculiarity index $\Delta a$
had to be defined as the difference between the individual $a$-values and those
of non-peculiar stars of the same colour. The locus of the $a_{\rm 0}$-values
has been called normality line

Virtually all peculiar objects with magnetic fields (CP2 and CP4 stars) 
have positive $\Delta a$ values up to +146\,mmag (El'Kin, Kudryavtsev, \& Romanyuk 2002). 
Only extreme cases of the CP1 and CP3 group 
exhibit marginally peculiar positive $\Delta a$ values (+20 mmag) whereas Be/Ae/shell and 
metal-weak (e.g. $\lambda$ Bootis group) stars exhibit significant negative ones 
(up to $-$35\,mmag). 
The index $(g_{\rm 1} - y)$ shows an excellent correlation with $(b - y)$ 
and can be used as indicator for the effective temperature (Kupka, Paunzen, \&
Maitzen 2003).

There are several important issues which can be answered from the
observational side by establishing the incidence of chemically peculiar
stars in the LMC. Most important is the question if there is an
influence of metallicity on the (non-)presence of peculiarities since
metallicity seems to distinctly influence the star formation scenario
in the sense that the formation of larger clusters is only possible
in low metallicity media. It seems worthwhile to investigate whether
the formation of magnetic peculiar and (non-magnetic) $\lambda$ Bootis stars occurs 
in the same proportion to ``normal'' stars for all degrees of metallicity. 
Furthermore, one should ask whether different general magnetic field 
strengths in the surrounding area of star formation will lead to the same frequency 
of magnetic stars. While one could try to find a relationship concerning 
metallicity also in our
own Galactic disc, the Magellanic Clouds offer a scenario with distinctly 
lower metallicity and lower magnetic field strength.
In order to explain CP stars stellar models are needed which take into
account different metallicities, ages and magnetic field strengths.

In this paper we present $\Delta a$ photometry for two representative fields
in the LMC. One field samples the LMC field population in the bulge whereas the other
one is centered on NGC 1711, a young globular cluster. In total, 1426 stars have
been observed from which 22 show significant positive whereas 9 exhibit negative
$\Delta a$ values. We derive the percentage of apparent chemically peculiar stars
and compare it with the numbers found for NGC 1866 and its surrounding as well as
the Milky Way.

\begin{table}
\caption{Observing log}
\begin{tabular}{ccccccccc}
\hline
        & Date & \#$_{N}$ & \#$_{g_{\rm 1}}$ & \#$_{g_{\rm 2}}$ & \#$_{y}$ \\
\hline
Field 1 & 23.11.1998 & 25 & $-$& 20 &  5 \\
        & 21.08.2001 & 28 & 20 & $-$ &  8 \\
		& 22.08.2001 &  4 &  4 & $-$ & $-$ \\
Field 2 & 05.01.2003 & 11 & 11 & $-$ & $-$ \\
	    & 06.01.2003 & 24 &  4 & 14 &  6 \\
        & 07.01.2003 & 10 & $-$& $-$ & 10 \\
\hline
\label{log}
\end{tabular}
\end{table}

\section{Target selection, observation and reduction process} \label{Sect2}

The first field is located in the bulge of the LMC on its eastern
edge (Cioni, Habing \& Israel 2000) centered at $\alpha$\,=\,04:45:20 and 
$\delta$\,=\,$-$69:12:00 (2000.0). There are no prominent
clusters or H{\sc ii} regions within this field in order to ensure that
only the field population of the LMC is observed.
The second field is centered on NGC 1711, a well studied
young globular cluster serving as a comparison to the published 
$\Delta a$ photometry of NGC 1866 and its surrounding (Maitzen
et al. 2001).

{\it NGC 1711}: The only recent detailed investigation, to our knowledge, of this 
cluster was done by Dirsch et al. (2000) who derived the following parameters within the
Str{\"o}mgren photometric system: log\,$t$\,=\,7.70(5), [Fe/H]\,=\,$-$0.57(17)\,dex. However,
these data are not available for the community (Sect. \ref{Sect3}). Therefore, we are
not able to testify their results. Kubiak (1990) lists an age of log\,$t$\,=\,7.3 on
the basis of solar abundant isochrones and Johnson $BVI$ photometry. Mateo (1988),
on the other hand, derived an age between log\,$t$\,=\,7.3 and 7.7, respectively.

\begin{figure*}
\plottwo{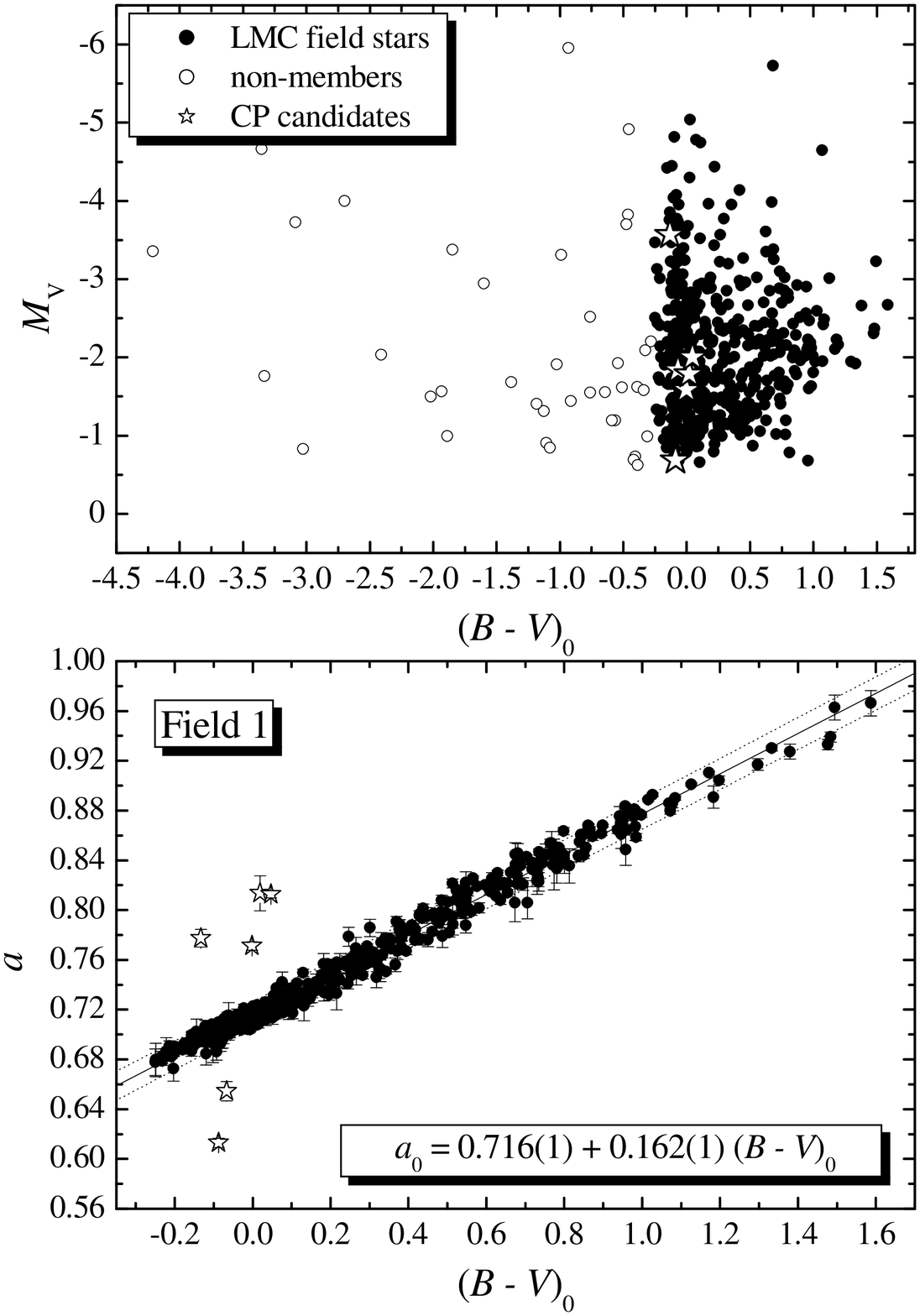}{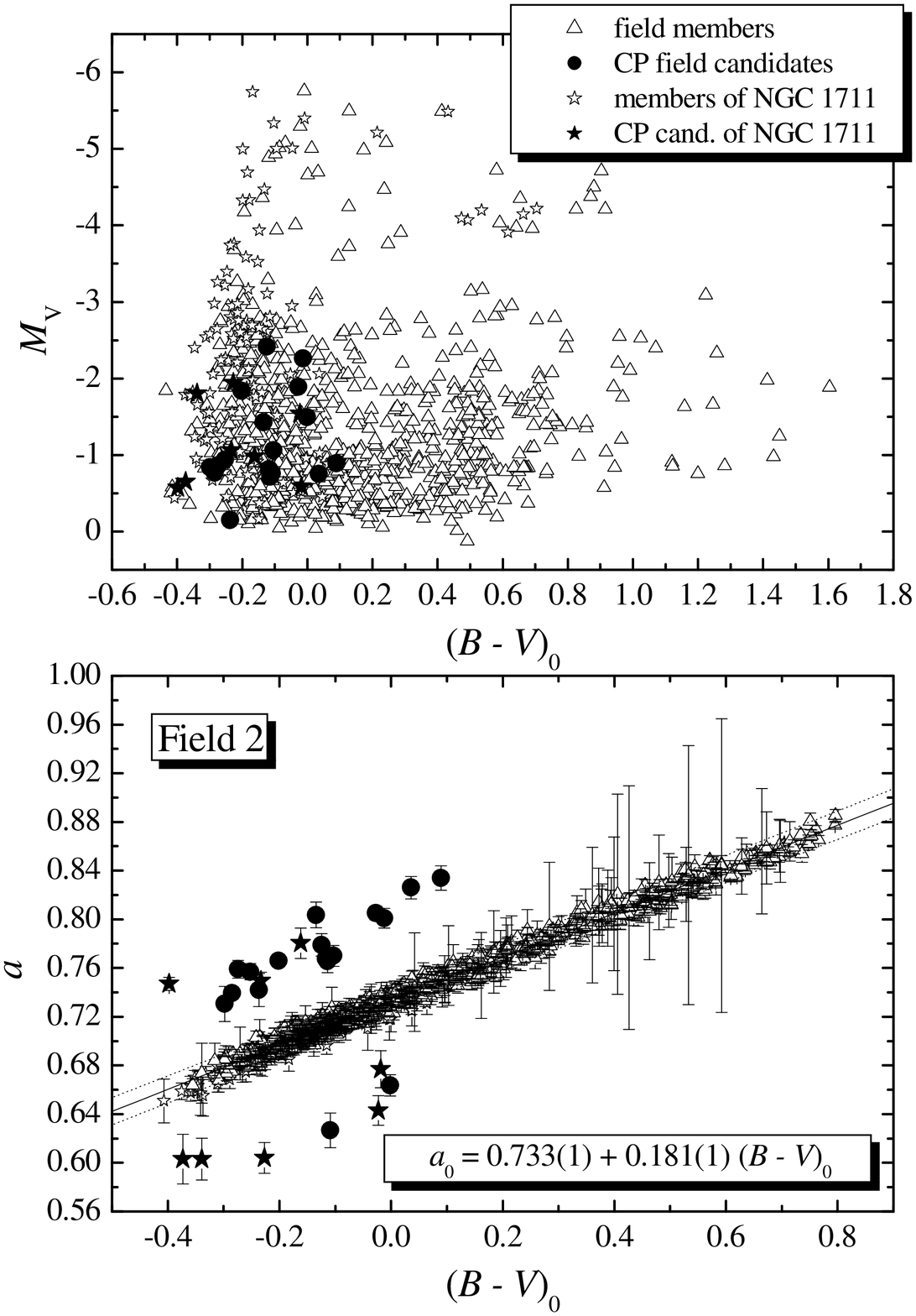}
\caption{Observed diagrams for our two program fields in the LMC.
The solid line is the
normality line whereas the dotted lines are the confidence intervals
corresponding to 99.9\%. The error bars for each individual object
are the mean errors. The measurement errors of $M_V$ are much smaller than the
symbols and have been omitted. The scales for the upper and lower diagrams 
are different because the relevant range for peculiar objects in the $a$ 
versus $(B-V)_0$ diagrams seemed worthwhile to be shown. The $M_V$ and 
$(B-V)_0$values were calculated using a distance modulus of 18.5 and a reddening
$E(B-V)\,=\,0.1$\,mag.}
\label{result}
\end{figure*}

{\it Field population}: it consists of a main sequence similar to that of NGC 1711
and an older population 
with a vertical extension of the red clump. Such a characteristics was observed
in several areas of the LMC (Dolphin 2000). It seems that stellar evolution was
rather uniform throughout the LMC until 200 Myr ago. Since then, a significant
increase can be inferred from observational data. The different populations 
are also seen in Fig. \ref{result} especially for the surrounding of NGC 1711. 

The photometric observations within the $\Delta a$ system
were performed at the Complejo Astron\'omico el Leoncito (CASLEO) 
using the 2.15m telescope (observer: O.I.~Pintado). The focal reducer
yields a scaling of 0.813$\arcsec$\,pixel$^{-1}$ and a field of view
of about 9.5$\arcmin$ using a TEK-1024 CCD. 

\begin{table}
\caption{Characteristics of the used filters.}
\begin{tabular}{lccccc}
\hline
Filter & $\lambda_{\rm C}$ & Bandwidth & Transmission \\
& [\AA] & [\AA] & [\%] \\
\hline
$g_{\rm 1}$ & 5027 & 222 & 66 \\
$g_{\rm 2}$ & 5205 & 107 & 50 \\
$y$ & 5509 & 120 & 54 \\
\hline
\label{filters}
\end{tabular}
\end{table}

The observations were performed in the years 1998, 2001 and 2003.
The detailed log is listed in Table \ref{log}. We have checked the
behaviour of the detector and the intrinsic transformations of all
data and found no significant trends or deviations from one year to
the other. The results for the open clusters in the Milky Way which
were observed during the same runs, have already been published by
Paunzen, Pintado \& Maitzen (2002b, 2003). This includes NGC 3114, a well 
studied open cluster with published photoelectric $\Delta a$ 
values which was used as a further test of our photometric 
system yielding high coincidence of both methods. 

The integration times were chosen between one to ten minutes in order
to measure the whole magnitude range within the linear region of
the CCD. In total, 102 frames in all filters for the two fields
in the LMC were obtained. The used filters are listed in Table
\ref{filters}. They have been working horses during the last
ten years.

The basic reductions (bias-subtraction, dark-correction, 
flat-fielding) were carried out within standard IRAF V2.12 routines.
A point-spread-function-fitting procedure within the
IRAF task DAOPHOT was performed, typically more than 30 individual,
single stars were used to derive the PSF for each frame.
Because of instrumentally induced offsets and different airmasses
between the single frames, 
photometric reduction of each frame was performed separately and the measurements were then 
averaged and weighted by their individual photometric error. This is justified because
the ``standard'' as well as program stars are on the same frame.

The table of all program stars with their photometric quantities and corresponding
errors are only available at SIMBAD or from the first author upon request.

\section{Calibration} \label{Sect3}

For both fields we have used the results of the extensive $UBVR$ CCD survey conducted by
Massey (2002) to calibrate and check our data. He presented accurate
measurements for approximately 250000 objects brighter than 18$^{th}$
magnitude in the Small and Large Magellanic Cloud. Although there
are several other sources of photometric data available (e.g. Holtzman
et al. 1997), only data of Massey (2002) were included in our analysis
in order to guarantee a sample free of bias on the basis of one 
widely accepted photometric standard system.

We have also tried to obtain the Str{\"o}mgren 
$uvby\beta$ photometry from Dirsch et al. (2000) 
published for NGC 1711 (and five other clusters in the LMC).
But it is neither available in electronic form nor upon request from the 
authors. 

For the calibration of our $y$ and $(g_1-y)$ to standard Johnson $V$ 
as well as $(B-V)$ values, we have chosen all stars in common with Massey (2002). 
This procedure gives a sample of 372 objects. A linear least square fit resulted in:
$$V\,=\,-2.67(9)\,+\,0.88(2)\cdot y$$ 
$$(B-V)\,=\,1.24(6)\,+\,1.44(8)\cdot (g_1-y)$$ 
In parenthesis are the errors in the final digits of the corresponding quantity.

Throughout this paper we use a distance modulus of 18.5 for the LMC taken from 
Alves (2004) who reviewed and summarized all relevant measurements as well
as their methods.

For the estimation of the reddening towards the observed fields, the maps
of Oestreicher, Gochermann \& Schmidt-Kaler (1995) were taken. Both fields are
located in areas with 0.05\,$<$\,$E(B-V)$\,$<$0.12\,mag. This is in line with the
results for NGC~1711 by Dirsch et al. (2000) who derived $E(B-V)$\,=\,0.09(5)\,mag
on the basis of Str{\"o}mgren $uvby$ photometry.
We have therefore adopted $E(B-V)$\,=\,0.1\,mag for both fields.
Our fields are not covered by the extensive survey
of Subramaniam (2005) who presented $E(V-I)$ values for 1123
locations in the LMC on the basis of the Optical Gravitational Lensing Experiment II.

We have checked the adopted reddening value and distance modulus by
applying appropriate isochrones (Claret, Paunzen \& Maitzen 2003, Claret
2005) to the
photometric measurements. Our results for NGC 1711 are in good agreement
with those from Kubiak (1990) and Dirsch et al. (2000).

For the normality lines, the $(g_1-y)$ measurements were converted into
$(B-V)$ values and dereddened. For Field 1 and 2 the following correlations
were found, respectively:
$$a_0\,=\,0.716(1)\,+\,0.162(1)\cdot (B-V)_0$$
$$a_0\,=\,0.733(1)\,+\,0.181(1)\cdot (B-V)_0$$
The results are shown graphically in Fig. \ref{result}. The 3$\sigma$ limit for
both fields is at $\pm$0.012\,mag. The non-members in Field 1 are objects which are
much too blue in respect to the main sequence. We can only speculate about their
nature since no independent $(B-V)$ values are available in the literature.
The most probable explanation is emission in the $g_1$ filter which would
yield such a blue color index. Some of these object might also be very distant
galaxies.

\begin{table}
\caption{Statistics of observed stars in the spectral range
up to F2 divided into Field 1 (centered at $\alpha$\,=\,04:45:20 and 
$\delta$\,=\,$-$69:12:0; 2000.0), NGC 1711 and Field 2 (surrounding
field population of NGC 1711). $N_{all}$ is the total number of
measured objects within the spectral range up to F2 or $(B-V)_0\,=\,0.3$\,mag, 
$N_{+}$ and $N_{-}$ are the numbers with significant
positive as well as negative $\Delta a$ values (Fig. \ref{result}, Table \ref{cps}). The lower
panel summarizes the results from Maitzen et al. (2001) who investigated the
spectral range from B8 to F2.}
\begin{tabular}{lcccccc}
\hline
Name & $N_{all}$ & $N_{+}$ & $N_{-}$ & $N_1$ & $N_2$ & $N_3$ \\
\hline
NGC 1711 & 109 & 3 & 5 & $-$ & 2.8 & 4.6 \\
Field 1 & 331 & 4 & 2 & 0.9 & 1.2 & 0.6 \\
Field 2 & 622 & 15 & 2 & 0.6 & 2.4 & 0.3 \\
\hline
NGC 1866 & 261 & 4 & 3 & 1.5 & $-$ & 1.1 \\
Field & 1239 & 4 & $-$ & 0.3 & $-$ & $-$ \\
\hline
\multicolumn{7}{l}{$N_1$: $100\cdot\frac{N_{+}}{N_{all}}$\,(B8 to F2), 
$N_2$: $100\cdot\frac{N_{+}}{N_{all}}$} \\
[-2mm] \\
\multicolumn{3}{l}{$N_3$: $100\cdot\frac{N_{-}}{N_{all}}$} \\
\label{percent}
\end{tabular}
\end{table}

\section{Analysis} \label{Sect4}

A summary of
the results from the CCD $\Delta a$ photometric measurements in the LMC is given
in Table \ref{percent}. In total, 2562
objects of which 30 exhibit significant positive whereas 12 have negative $\Delta a$
values (Table \ref{cps}). The latter are either Be/shell or metal-weak objects 
and have, in general, no strong magnetic fields.

The only paper about the number of apparent magnetic CP stars in the LMC was 
published by Maitzen et al. (2001) who found an overall percentage of CP stars 
for NGC 1866 (log\,$t$\,=\,8.0, [Fe/H]\,=\,$-$0.43(18)\,dex; Hilker, Richtler \& Gieren 1995)
of 1.5\% whereas the incidence within the LMC field was decisively 
less (0.3\%). They have investigated a spectral range from B8 to F2 or 
$-0.1\,<\,(B-V)_0\,<0.3$\,mag using the same observational technique and 
instrument as in this work. We have included their results in Table \ref{percent}.

It has to be emphasized that Kupka et al. (2003) showed on the basis of a
synthetic $\Delta a$ photometric system that only a shift of the normality line
by about $-3$~mmag assuming an average metallicity of $[Fe/H]\,=\,-0.5$\,dex 
relative to those in the solar neighborhood occurs. The absolute $\Delta a$ values for
CP stars are not affected.

{\it Objects with significant negative $\Delta a$ values}: this group consists of 
either classical emission line or metal-weak objects (like $\lambda$ Bootis
stars). In total, nine of such objects were detected.
Without any further photometric or spectroscopic data, no clear decision 
about the true nature can be drawn. However, it is not surprising to find a rather
significant amount of these objects in NGC 1711 since it is a very young cluster
still connected with stellar activity and emission of all kinds. Metal-weak main
sequence stars in the relevant spectral range (the so-called $\lambda$ Bootis group)
are very scarce (compared to metal-strong objects) in the field population of 
the solar neighborhood (only a maximum of 2\% of all stars) and almost absent 
in open clusters older than log\,$t$\,=\,7.30 in our
Milky Way (Gray \& Corbally 2002). This might be caused by accretion
from a diffuse interstellar cloud together with diffusion as most probable 
mechanism for the $\lambda$ Bootis phenomenon (Kamp \& Paunzen 2002). 
This scenario works at any stage of stellar evolution as soon as the star passes a diffuse 
interstellar cloud. Its dust grains are blown away by the stellar
radiation pressure, while the depleted cloud gas is accreted onto the star. 
This would naturally generate an abundance pattern as found for the $\lambda$ Bootis
group, namely surface underabundances of most
Fe-peak elements and solar abundances of lighter elements (Carbon, Nitrogen, Oxygen, and 
Sulphur). Since
denser clouds within clusters dissipate on time scales of about 50 Myr, no $\lambda$ Bootis
stars are expected in older associations which make their appearance in clusters
of the LMC rather unlikely.   

{\it The incidence of Be/Ae/shell as well as metal-weak stars in the LMC}: these are the
only groups which exhibit significant negative $\Delta a$ values. The Be/shell 
population in the LMC is well studied (Keller, Wood \& Bessell 1999). They found a
fraction of main sequence stars that are Be objects between 0.1 and 0.34 
(mean value of the Milky Way is about 0.17) varying for
six very young clusters and their field population in the LMC. No correlation of
the incidence with the age or metallicity was detected. Gray \& Corbally (2002)
concluded that the frequency of $\lambda$ Bootis stars in the Galactic field
among main sequence A to F objects is about 2\% (no investigation for the LMC 
is available up to now). The age distribution of both 
groups are rather different. Whereas Be/Ae/shell stars are preferably found at young
ages with log\,$t$\,$<$\,7.5, the members of the $\lambda$ Bootis group in the Milky Way 
are evolved with a peak at log\,$t$\,=\,9.0 (Paunzen et al. 2002a). The detection capability
of the $\Delta a$ photometric system for a 3$\sigma$ limit of 0.012\,mag is at 50\%
as well as 10\% for the $\lambda$ Bootis and Be/shell group, respectively 
(Paunzen, St{\"u}tz \& Maitzen 2005). This means that 10\% of the 
complete Be/Ae/shell as well as 50\% of the $\lambda$ Bootis population can 
still be detected with a high statistical significance at a limit of 0.012\,mag. 
If we take the maximum percentage of the relevant group members from 
Keller et al. (1999) and Gray \& Corbally (2002), an upper frequency of 
4.4\% of objects with negative $\Delta a$ values (3.4\% Be/Ae/shell as well as 1\% 
$\lambda$ Bootis) among all main sequence B to F type stars should 
occur in the LMC for the derived detection limit of 0.012\,mag. 
The number of such objects in NGC 1711 is 4.6\% and 
significantly less in the other fields. The presence of these stars is, in general,
depending on denser interstellar clouds ($\lambda$ Bootis group) as well as stellar
activity which is closely correlated with the age. However, it showes that $\Delta a$
photometry, taking into account a certain
detection limit, is able to find essentially the same frequency of Be/Ae/shell 
and metal-weak stars 
as other photometric as well as spectroscopic surveys. 

{\it Objects with significant positive $\Delta a$ values}: the detected 22 objects
are most certainly magnetic CP (CP2 and/or CP4) stars.
The main characteristics of the classical CP2 stars are: peculiar
and often variable line strengths, quadrature of line variability
with radial velocity changes, photometric variability with the same
periodicity and coincidence of extrema in the presence of slow rotation. 
Overabundances of several orders
of magnitude compared to the Sun were derived for heavy elements
such as Silicon, Chromium, Strontium and Europium (Bagnulo et al. 2002).
CP2 stars are found over the spectral range from early B to early F.
The CP4 stars comprise 
helium weak B type objects. As the CP2 group, they have strong magnetic fields, 
elemental surface inhomogeneities together with photometric variations. Several 
objects also show emission in the optical spectral range and signs of mass loss 
(Wahlgren \& Hubrig 2004). The peculiar objects found here should be a mixture of 
both groups. Only for the
two objects cooler than A0, $(B-V)_0\,=\,0$\,mag, i.e. we can assume a membership to 
the CP2 group.

\begin{table}
{\scriptsize
\caption{Peculiar objects found during our survey in the LMC. The zero point for Field
1 is star No. 22207 (Massey 2002; $\alpha$\,=\,04:54:49.84,  
$\delta$\,=\,$-$69:11:26.4, 2000.0) as well as No. 7391 ($\alpha$\,=\,04:50:17.20, 
$\delta$\,=\,$-$69:57:57.3, 2000.0) for Field 2 (including NGC 1711), respectively.}
\begin{tabular}{rrrlll}
\hline
No. & \multicolumn{1}{c}{X} & \multicolumn{1}{c}{Y} & \multicolumn{1}{c}{$(B-V)_0$} & 
\multicolumn{1}{c}{$\Delta a$} & \multicolumn{1}{c}{$V$} \\
    & \multicolumn{1}{c}{[Pixel]} & \multicolumn{1}{c}{[Pixel]} & \multicolumn{1}{c}{[mag]} & 
	\multicolumn{1}{c}{[mag]} & \multicolumn{1}{c}{[mag]} \\        
\hline
Field 1 \\
     35 & $-$52.2 & $-$54.0 & +0.019(9) & +0.094 & 17.024(1) \\
	 40 & $-$47.8 & +107.9 & $-$0.087(4) & $-$0.089 & 18.125(5) \\
	118 & +20.8 & +220.3 & $-$0.067(7) & $-$0.050 & 16.973(9) \\
	250 & +148.6 & +359.1 & $-$0.133(8) & +0.083 & 15.241(9) \\
	278 & +175.5 & $-$6.5 & $-$0.001(3) & +0.055 & 17.227(3) \\
	483 & +327.9 & $-$78.5 & +0.047(3) & +0.089 & 16.807(3) \\ 
	\hline
Field 2 \\
    105	& $-$307.4 & +206.3	& $-$0.284(3) & +0.058 & 18.041(2) \\
    107	& $-$305.0 & $-$96.5 & $-$0.109(8) & $-$0.086 & 18.052(8) \\
    117	& $-$300.7 & +212.6	& $-$0.125(5) & +0.069 & 16.397(5) \\
    186	& $-$269.2 & +264.8	& $-$0.298(8) & +0.052 & 17.970(3) \\
    196	& $-$265.8 & +189.3	& $-$0.012(4) & +0.071 & 16.551(5) \\
    199	& $-$265.2 & +199.7	& $-$0.274(4) & +0.076 & 17.947(4) \\
    202	& $-$263.8 & +257.8	& $-$0.201(2) & +0.070 & 16.972(2) \\
    232	& $-$249.8 & +262.7	& $-$0.252(3) & +0.070 & 17.875(3) \\
    668	& $-$102.5 & +229.6	& +0.036(5) & +0.087 & 18.060(5) \\
    708	& $-$89.8 & +139.8 & $-$0.103(4) & +0.056 & 17.750(4) \\
    789	& $-$56.2 & +330.5 & $-$0.237(7) & +0.052 & 18.661(3) \\
    825	& $-$43.5 & +289.3 & $-$0.114(5) & +0.054 & 18.096(5) \\
    911	& +10.1 & +230.0 & $-$0.117(3) & +0.056 & 17.995(3) \\
    936	& +31.4 & +95.9 & $-$0.027(3) & +0.078 & 16.919(3) \\
    951	& +40.2 & +243.5 & +0.090(5) & +0.085 & 17.918(6) \\
    982	& +73.5 & +264.4 & $-$0.134(6) & +0.095 & 17.385(3) \\
    998	& +98.6 & $-$12.5 & $-$0.001(2) & $-$0.069 & 17.314(1) \\
NGC 1711 \\
    390	& $-$186.9 & +100.0	& $-$0.398(2) & +0.087 & 18.238(2) \\
    481	& $-$158.0 & +83.6	& $-$0.227(6) & $-$0.087 & 16.871(7) \\
    499	& $-$151.3 & +75.9	& $-$0.162(3) & +0.077 & 17.834(3) \\
    554	& $-$137.8 & +104.5	& $-$0.339(9) & $-$0.068 & 17.009(9) \\
    572	& $-$132.6 & +47.8	& $-$0.022(7) & $-$0.085 & 17.271(6) \\
    586	& $-$129.8 & +126.6	& $-$0.233(2) & +0.059 & 17.751(1) \\
    652	& $-$106.3 & +55.4	& $-$0.373(9) & $-$0.062 & 18.159(9) \\
    719	& $-$86.4 & +66.7	& $-$0.018(8) & $-$0.052 & 18.222(6) \\ 
\hline
\label{cps}
\end{tabular}
}
\end{table}

{\it The incidence of magnetic CP stars in the Milky Way}: two aspects have
to be taken into account when comparing the number of CP stars in the Milky
Way and the LMC. First of all, the metallicities in the solar neighborhood
and the galactic open clusters are significantly higher than in the young 
associations in the LMC which hardly reach $-$0.5\,dex. The second issue is about
the evolutionary effect on the incidence of magnetic CP stars. 
The Hipparcos data suggest that the CP group behave just like apparently normal 
stars in the 
same range of spectral types occupying the whole width of the main sequence with
kinematic characteristics typical of thin disk stars younger than about 1\,Gyr
(G{\'o}mez et al. 1998). This finding was supported by P{\"o}hnl, Maitzen \& Paunzen
(2003) who investigated four young open clusters (ages between 10 an 140\,Myr) 
with known CP2 members establishing the occurrence of such objects at very
early stages of the stellar evolution. However, Hubrig, North \& Mathys (2000) challenged these
results and claim that the distribution of the magnetic CP stars of masses below 
3\,M$_{\odot}$ in the Hertzsprung-Russell-diagram differs from that of the normal 
stars in the same temperature range at a high level of significance, magnetic stars being 
concentrated toward the center of the main sequence band. In particular, they argue that 
magnetic fields are detected only in stars which have already completed at least 
approximately 30\% of their main sequence life time. This somewhat discrepant result 
might be understood by the detectability of resolved Zeeman 
patterns which requires a specifically slow rotation. This gives preference to finding such 
objects in advanced phases on the main sequence band where rotational velocities have 
been decreased by the growth of stellar radii. \\
Abt (1979) investigated the dependence
of the percentage of CP members in open clusters on their age. His sample,
divided into magnetic Ap (Si) and Ap (Sr,Cr) stars, includes data
for 14 galactic open clusters as well as associations, a rather small number
if one takes into account that at least 1500 open clusters are known
in our Milky Way (Dias et al. 2002). The frequency of Ap (Si)
objects with $-$1.3\,$<$\,$M_V$\,$<$1.4\,mag increases from 4\% to 8\% during 
log\,$t$\,=\,7.0 and 8.0, respectively, whereas the youngest Ap (Sr,Cr)
stars are found at log\,$t$\,=\,7.5 (3\% at log\,$t$\,=\,8.0). For, at that
time, known peculiar stars (Osawa 1965)
of the Bright Star Catalogue, Abt (1979) gives a number of 6.5\% for the
same magnitude and declination range as the investigated members of open clusters. No value for 
the CP4 group is listed in Abt (1979).
Another source for a statistically analysis of galactic CP field stars is 
the conference paper by Schneider (1993) who considered also the Supplement to the
Bright Star Catalogue and the CP stars listed in Renson (1991). He lists for the
magnitude range $V$\,$<$\,7.1\,mag a lower limit of 
16\% (CP2 and CP4) and 5\% (CP2 only) for stars hotter and cooler than A0, respectively.
These numbers should be taken with care since Schneider (1993) quotes that ``the single
values should not be taken too serious because the MK classification influences the
single values strongly''. From the mentioned sources we conclude that one would expect
an incidence for magnetic CP stars of at least 6\% over the relevant spectral range 
up to F2 main sequence objects. This number has to be treated as statistically lower limit
for the percentage of magnetic CP stars for any test sample of main sequence objects in
the Milky Way. The only limitation might be very young open clusters and associations
(log\,$t$\,$<$\,7.0) where the incidence seems slightly less than 6\% 
(Borra, Joncas \& Wizinowich 1982).

{\it The incidence of magnetic CP stars in the LMC}: first of all, one has to
correct the observed ratio of CP stars for loss of objects because of the detection limit.
Paunzen et al. (2005) estimated a detection rate of 90\% for CP2 and CP4 objects
for a limit of 0.012\,mag. Incorporating this fact for the values listed in Table 
\ref{percent}, a maximum value of about 2.5\% (the percentage for NGC 1866 might be
up to 3\% if we extrapolate the result) for the incidence of magnetic CP stars in 
the whole spectral range up to F2 is inferred. This number is based on the 
measurements of 2562 individual
objects. The occurrence of magnetic CP stars in the LMC is therefore only about half the
value as in the Milky Way.

\section{Conclusions and outlook}

We have continued our search for CP stars of the upper main sequence in the LMC 
applying the tool of $\Delta a$ photometry which measures the 
characteristic broad band absorption feature located around 5200\AA.
This flux depression is most certainly a consequence 
of the non-solar elemental abundance of CP and related 
objects in the presence of a strong stellar magnetic field. 
Our final goal is to establish the
incidence of those objects and to compare it with the values found for the Milky Way.
Such an investigation is very important for stellar evolution theory since the
overall metallicity of the LMC is reduced by up to 0.5\,dex compared to the Sun. 
Furthermore, its
global magnetic field consists of a coherent axisymmetric spiral of field 
strength which is weaker than that of the Milky Way. Both parameters play
a key role 
in the context magnetic CP stars because the origin of their magnetic fields is still a 
matter of debate: those who favor the survival of frozen-in fossil fields originating 
from the medium out of which the stars were formed are in opposition to those 
following the idea that a dynamo mechanism is acting in the stellar interior. 

In total, 2562 objects in two young globular clusters and three different 
field populations in the relevant spectral range from early B to early F 
have been photometrically investigated. From this sample, 30
objects (most certainly magnetic CP stars) exhibit significant positive 
whereas 12 have negative (metal-weak or Be/Ae/shell stars) $\Delta a$ values.

From a comparison of the published statistics of magnetic CP stars in the
galactic field and open clusters in the Milky Way, we concluded that the incidence
in the LMC is only about half the value from the Milky Way. This is already a very
strict observational constraint for the standard evolutionary model. Our derived number of 
metal-weak and Be/Ae/shell stars in the observed fields match excellent the numbers
derived from spectroscopic investigations in other fields of the LMC.

As next ambitious step, we will try to establish the incidence of CP stars in
the Small Magellanic Cloud. A galaxy with an ever lower metallicity than the LMC.
Furthermore, high resolution spectroscopy of the detected CP stars is needed
to compare their detailed elemental abundances with those of galactic 
counterparts. 

\section*{Acknowledgments}

The CCD and data acquisition system at CASLEO has been partly financed
by R.M. Rich through U.S. NSF Grant AST-90-15827.
This research was performed within the projects  
P17580 and P17920 of the Austrian Fonds zur F{\"o}rderung der 
wissen\-schaft\-lichen Forschung (FwF) and the City of
Vienna Hochschuljubil{\"a}umsstiftung project: $\Delta a$ Photometrie in der 
Milchstrasse und den Magellanschen Wolken, H-1123/2002.
Use was made of the SIMBAD database, operated at CDS, Strasbourg, France
and NASA's Astrophysics Data System.

\label{lastpage}

\end{document}